\documentclass[journal]{IEEEtran}
\usepackage{caption}
\usepackage{subcaption}
\usepackage{url}
\usepackage{bibentry}  
\usepackage{cite}
\usepackage{verbatim}
\usepackage{grffile}
\usepackage{algorithm}
\usepackage{algpseudocode}
\usepackage{tabularx}
\usepackage{xcolor,soul,framed} 
\colorlet{shadecolor}{yellow}
\usepackage{amsmath,amssymb,amsfonts}
\usepackage{graphicx}
\usepackage{bm}
\usepackage{textcomp}
\usepackage{mathtools}

\usepackage{mathrsfs}
\usepackage{enumitem}
\usepackage{float}
\usepackage{multirow}
\usepackage{multicol}
\usepackage{array}
\usepackage{hyperref}
\usepackage{cancel}
\usepackage{titlesec}
\usepackage{bbm}
\usepackage{tikz}
\usetikzlibrary{shapes,calc}
\usepackage{anyfontsize}
\usepackage{booktabs}
\usepackage{siunitx}

\begin{document}

\setlength{\abovedisplayskip}{2pt}
\setlength{\belowdisplayskip}{2pt}

\title{Traffic-Aware Cost-Optimized Fronthaul Planning for Ultra-Dense Networks}

\author{Anas S. Mohammed, Hussein A. Ammar, \IEEEmembership{Member, IEEE}, Krishnendu S. Tharakan, \IEEEmembership{Member, IEEE}, Hesham ElSawy, \IEEEmembership{Senior Member, IEEE}, 
and Hossam S. Hassanein, \IEEEmembership{Fellow, IEEE}
\thanks{This work is supported in part by NSERC DG number RGPIN-2023-03743.}
\thanks{A. S. Mohammed is with the Department of Electrical and Computer Engineering, Queen’s University, Kingston, Canada. E-mail: anas.m@queensu.ca.} 
\thanks{H. A. Ammar is with the Department of Electrical and Computer Engineering, Royal Military College of Canada, Kingston, Canada. E-mail: hussein.ammar@rmc.ca.}
\thanks{K. S. Tharakan, H. ElSawy and H. S. Hassanein are with the School of Computing, Queen’s University, Kingston, Canada. E-mail: \{k.tharakan, hesham.elsawy\}@queensu.ca, hossam@cs.queensu.ca.} \vspace{-30pt}}
{}
\maketitle
\begin{abstract}
The cost and limited capacity of fronthaul links pose significant challenges for the deployment of ultra-dense networks (UDNs), specifically for cell-free massive MIMO systems. Hence, cost-effective planning of reliable fronthaul networks is crucial for the future deployment of UDNs. We propose an optimization framework for traffic-aware hybrid fronthaul network planning, aimed at minimizing total costs through a mixed-integer linear program (MILP) that considers fiber optics and mmWave, along with optimizing key performance metrics. The results demonstrate superiority of the proposed approach, highlighting the cost-effectiveness and performance advantages when compared to different deployment schemes. Moreover, our results also reveal different trends that are critical for Service Providers (SPs) during the fronthaul planning phase of future-proof networks that can adapt to evolving traffic demands.
\end{abstract}
\vspace{-3pt}
\vspace{-1.1em}
\section{Introduction}\label{sec:introduction}
\vspace{-5pt}
\IEEEPARstart{T}{he} evolution of mobile networks has consistently brought substantial advancements with each generation, culminating in the capabilities of 5G. Nevertheless, demands for connectivity and data rates are rapidly intensifying as we move towards 6G and beyond, leading to ultra-dense networks (UDNs)~\cite{UDN}. However, the effective deployment of UDNs requires cooperation among Access Points (APs) to manage interference. This necessity has given rise to advanced network schemes such as cell-free massive MIMO (CF-mMIMO), which provides ubiquitous spatial coverage and high data rates through dense AP deployment~\cite{Book}. Despite these advantages, the widespread deployment of UDNs is primarily obstructed by the sole reliance on wired fronthaul technologies, such as fiber optics, to connect central processing units (CPUs) with APs, leading to significant infrastructural costs and scalability limitations~\cite{NO.1-Sources, FSOandFiberCFMIMO}. Conversely, utilizing wireless technologies to design a fronthaul network, such as mmWave or free-space optics (FSO), represent a promising alternative to wired counterparts by offering adequate capacities at lower deployment costs~\cite{UDN, NO.1-Sources, mmwave_source, FSOandFiberCFMIMO, nlosss}. Thus, planning fronthaul networks that are scalable, cost-effective, reliable and capable of supporting high data traffic is critical for the future deployment of UDNs.

To address these challenges, the motive of this letter is to investigate the feasibility and cost-effectiveness of an optimized hybrid fronthaul network that comprises fiber and mmWave technologies, to support the future deployment of UDNs. Our objective is to minimize the total cost of ownership (TCO) while achieving sufficient network performance against several benchmarks and across key metrics such as link and network capacities, and network reliability. Our results show that relying exclusively on a single fronthaul technology—whether fiber, or mmWave—may prove neither cost-effective nor resilient enough for future UDNs.

\vspace{-0.7em}
\section{Hybrid Fronthaul System Model}\label{Sec2}
\vspace{-3pt}
To align with Open Radio Access Networks (O-RAN) terminologies, we will refer to the CPU as the Distributed Unit (DU) from this point onward~\cite{O-RAN}. The fronthaul network investigated in this paper comprises of APs, DUs, and fronthaul links utilizing either fiber or mmWave. Without loss of generality, we consider a 2D square area of size $R \times R \; \text{m}^2 $ containing $L$ uniformly distributed APs with coordinates $ (x_{\ell}, y_{\ell}) \sim \mathcal{U}(0, R) $ for $ \ell = 1, 2, .., L $ in multiple realizations to account for actual deployment perturbations. Additionally, $W$ DUs are deployed with their placement optimized using K-means clustering, to ensure proper AP-DU associations based on proximity, where each AP $\ell$ is only associated with a single DU~\cite{NO.1-Sources}. The set of all APs is denoted by $\mathcal{L}$, and the set of all DUs is $\mathcal{W}$, where the cluster of all APs associated with DU $ w $ is represented by the subset $ \mathbf{L}_w \subseteq \mathcal{L}$. We assume point-to-point (P2P) links between APs and DUs using fiber, with constant capacity $R_{w\ell}^{\text{Fiber}}$. Key equipment typically include fiber cables and optical network units (ONUs) integrated with each AP $\ell$, and the optical transport unit (OTN) colocated at each DU $w$, comprising splitters, multiplexers (MUXs), optical line terminals (OLTs), and other equipment that manage optical signals from multiple ONUs and aggregate traffic at the DUs. While APs that use mmWave are equipped with $N_{\text{AP}}$ antennas, and DUs that serve mmWave-based APs are equipped with $N_{\text{DU}}$ antennas.

\vspace{-1em}
\subsection{Fronthaul Capacity Thresholds for APs}
\begin{figure}[t]
    \centering
    \vspace{-0.8em}
    \begin{subfigure}[b]{0.47\columnwidth}
        \centering
        \includegraphics[width=\textwidth, height=1.37in]{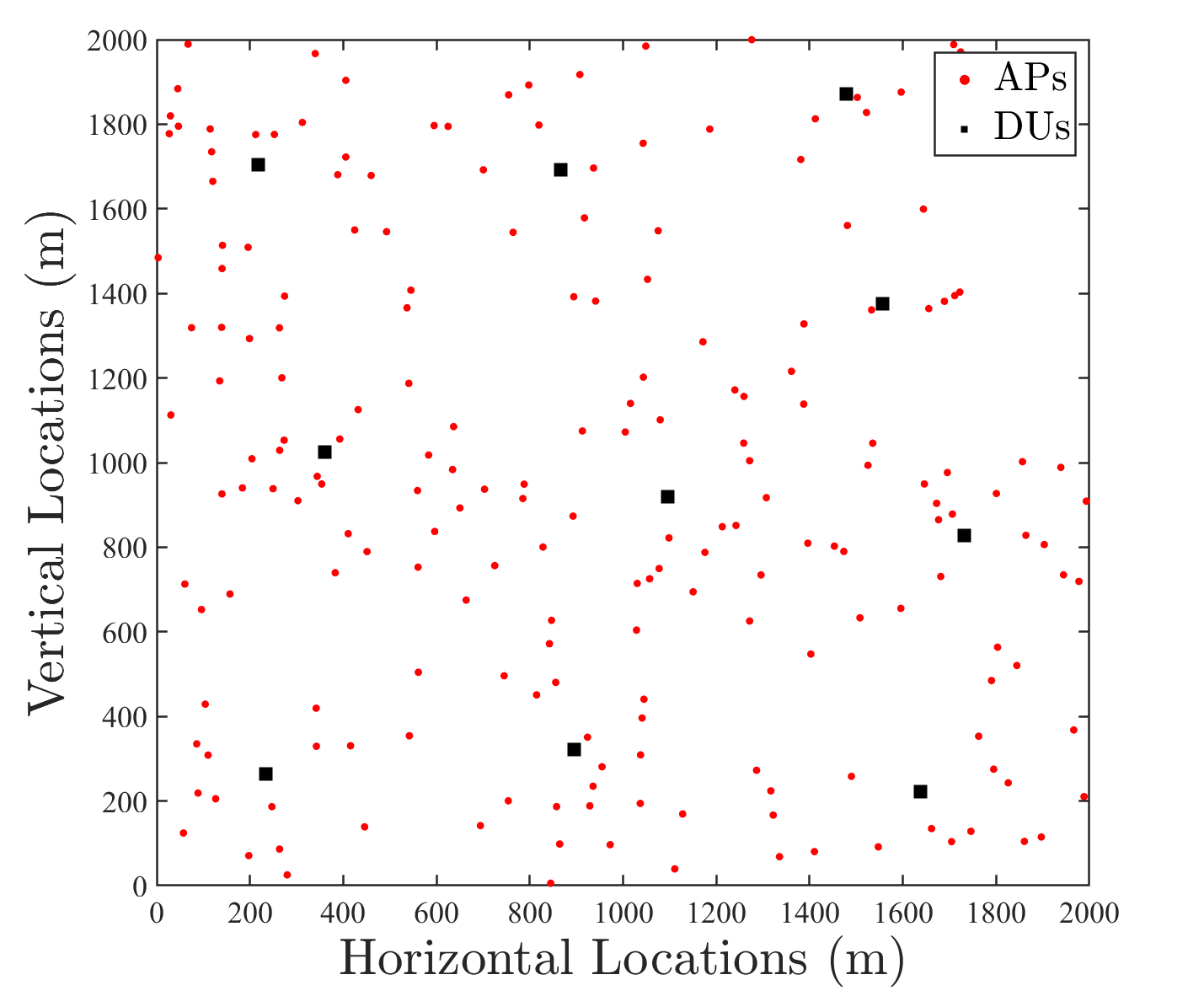}
        \caption{Optimized DUs placement.}
        \label{fig1.1}
    \end{subfigure}
    \hfill
    \begin{subfigure}[b]{0.49\columnwidth}
        \centering
        \includegraphics[width=\textwidth, height=1.37in]{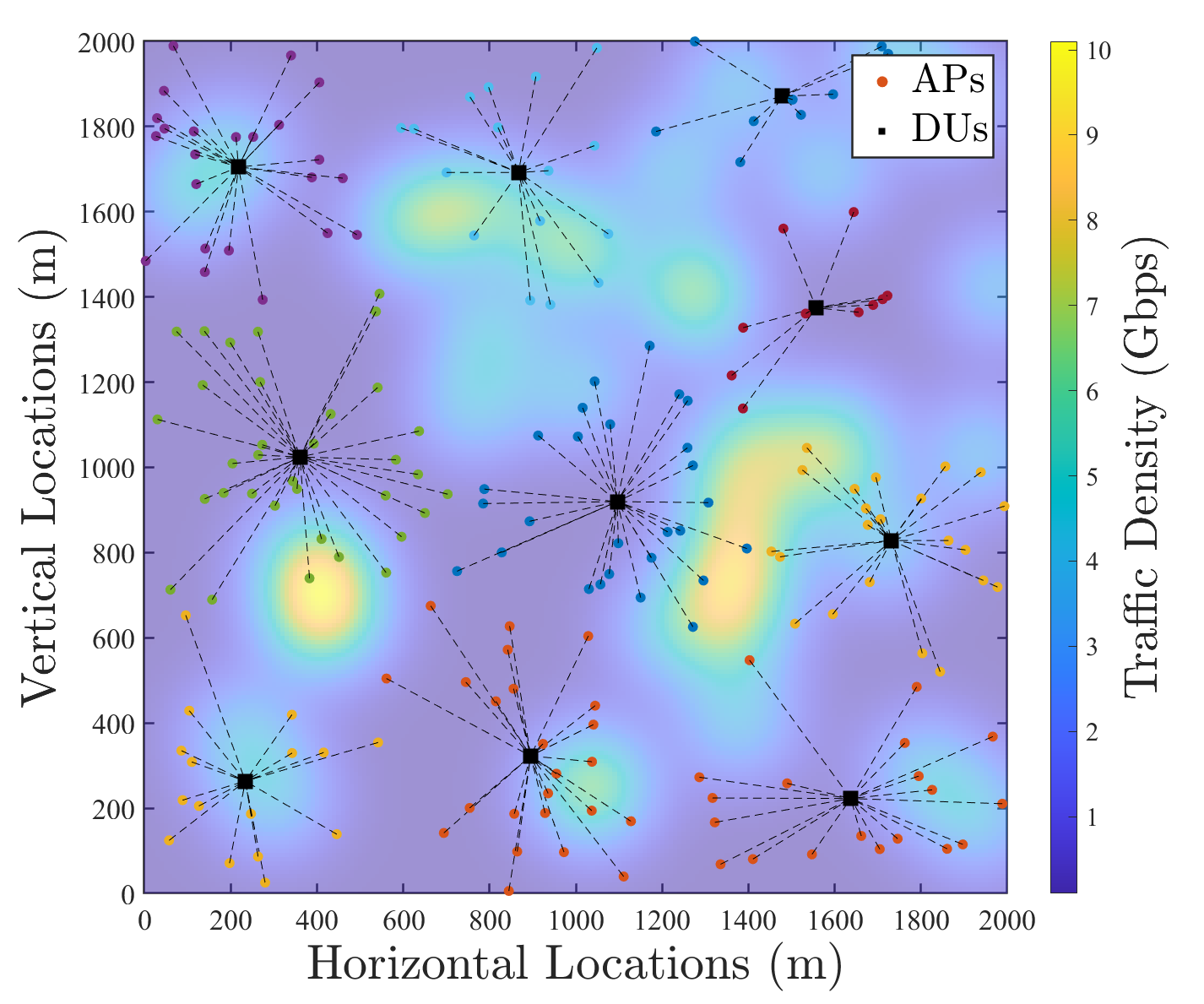}
        \caption{Combined network and traffic.}
        \label{fig1.2}
    \end{subfigure}
    \vspace{-0.4em}
    \caption{Sample network realization with optimized DUs placement and APs traffic thresholds assignment.}
    \label{fig1}
    \vspace{-2.5em}
\end{figure}

\vspace{-3pt}
After deploying APs and DUs across the network as shown from Figure \ref{fig1.1}, a traffic-aware capacity threshold $\psi_{\ell}$ is assigned to each AP $\ell$ based on the underlying traffic distribution utilizing the concept of hotspots, and modeled using a Gaussian distribution to represent the areas of peak demand within the network. We consider $N_s$ randomly distributed hotspots with coordinates $(x_s, y_s) \sim \mathcal{U}(0, \sigma_{s}^{2})$, where $\sigma_s$ controls the spread of hotspots. The normalized traffic distribution function, averaged over all hotspots, is given by~\cite{Generating-Traffic}:
\begin{equation}
      f_{\text{dist}}^{\text{norm}}(x, y) = \frac{1}{N_s} \sum_{s=1}^{N_s} \frac{1}{2\pi\sigma_{s}^{2}} e^{\left(-\frac{(x - x_s)^2 + (y - y_s)^2}{2\sigma_{s}^{2}}\right)}.
\end{equation}
$f_{\text{dist}}^{\text{norm}}(x, y)$ is then scaled to reflect realistic traffic values. This ensures that the peak traffic demand does not exceed a specified limit, $\mathbf{X_{}^{\text{max}}}$, and an offset is added to ensure the minimum traffic threshold is not below $\mathbf{X}_{}^{\text{min}}$.

\begin{equation}
    f_{\text{dist}}^{\text{scale}}(x, y) = \left( \frac{\mathbf{X^{\text{max}}}}{\max_{x, y} \{f_{\text{dist}}^{\text{norm}}(x, y)\}} \right)f_{\text{dist}}^{\text{norm}}(x, y),
\end{equation}
\begin{equation}
    f_{\text{traf.}}(x, y) = f_{\text{dist}}^{\text{scale}}(x, y) + \left(\mathbf{X}_{}^{\text{min}} -  \min_{x, y} \{f_{\text{dist}}^{\text{scale}}(x, y)\} \right).
\end{equation}
The traffic-aware fronthaul capacity threshold is assigned for each AP $\ell$ based on its position on the heatmap within the coverage region as seen in Figure \ref{fig1.2}, and as follows:
\begin{equation}
    \psi_{\ell} = f_{\text{traf.}}(x_{\ell}, y_{\ell}), \;\;\;\; \forall \ell \in \mathcal{L}.
\end{equation}
\vspace{-15pt}
\vspace{-1em}

\subsection{Channel Modeling for mmWave-based Fronthaul Links}
\vspace{-2.5pt}
For mmWave, we adopt 3GPP 38.901 Urban Microcell street canyon (UMi-SC) model to characterize both Line-of-Sight (LoS) and Non-Line-of-Sight (NLoS) propagation~\cite{3gpp}. In contrast to satellite and aerial-integrated networks, where LoS is not always guaranteed due to mobility~\cite{nlosss}, a typical approach that SPs adopt in terrestrial networks is to leverage the static nature of APs and DUs to ensure LoS during wireless fronthaul planning. Thus, LoS parameters are distance-based, while NLoS parameters, including the number of paths \(P \sim \mathcal{U}[1, 6]\) and angle-of-departure \(\theta_p \sim \mathcal{U}[-\frac{\pi}{2}, \frac{\pi}{2}]\), are randomly selected based on discrete uniform distribution to account for possible reflections, with small-scale fading modeled as complex Gaussian of zero mean and unit variance~\cite{mmwave_source}.
\begin{equation}
\text{PL}_{\ell}^{\text{LoS}} = 32.4 + 21 \log_{10}(d_{w \ell}) + 20 \log_{10}\left(f_{\text{c}}\right) + \mathcal{S}, 
\end{equation}
\begin{equation}
\text{PL}_{\ell}^{\text{NLoS}} = 32.4 + 31.9 \log_{10}(d_{w\ell}) + 20 \log_{10}\left(f_{\text{c}}\right) + \mathcal{S},
\end{equation}
where $f_{\text{c}}$ is the carrier frequency in GHz, and $\mathcal{S} \in \mathcal{N}(0, \sigma^2)$ is the shadowing with standard deviation 4 dB and 8.2 dB for LoS and NLoS, respectively~\cite{3gpp}. In downlink (DL), the signal is transmitted from a DU to its associated APs, and zero-forcing (ZF) analog beamforming is employed with the precoding matrix for any DU $w$ defined as $\mathbf{F}_w \triangleq [{\bf f}_{w1} \ldots {\bf f}_{w L_w}] \in \mathbb{C}^{N_{\text{DU}} \times N_{\text{AP}}}$, where ${\bf f}_{w\ell} \in \mathbb{C}^{N_{\text{DU}}}$ is the beamforming vector used to serve AP $\ell$. We assume a network of quantized phase shifters to satisfy the practical constraints of mmWave. Hence, the beamforming vector ${\bf f}_{w\ell}$ is selected from a finite set \(\mathcal{F}_w\). For a phase shifter with $q$ quantization bits, we have $2^q$ phase shift values defined by $\mathcal{Q} = \{0, \frac{\pi}{2^q}, \ldots, \frac{(2^q - 1)\pi}{2^q}\}$, and the set of all possible beamforming vectors for DU $w$ is~\cite{mmwave_source}:
\begin{equation}
    \mathcal{F}_w = \left\{ \frac{\begin{bmatrix} e^{j\phi_1} ... e^{j\phi_{N_{\text{DU}}}}  \end{bmatrix}^T}{N_{\text{DU}}} ;  \phi_i \in \mathcal{Q}, \forall i \in \{1, .., N_{\text{DU}}\}
   \right\},
\end{equation}
and the signal received at AP $\ell$ is given as:
\begin{equation}
\mathbf{r}_{\ell}^{\text{mmW}} =  \sqrt{p_{t}^{\text{mmW}}} \mathbf{h}^T_{w \ell}\mathbf{f}_{w \ell} \varsigma_{\ell} + \!\! \underbrace{\sum_{\substack{w' = 1, \\ w' \neq w}}^{W} \sqrt{p_{t}^{\text{mmW}}} \mathbf{h}_{w' i}^T \mathbf{f}_{w' i}\varsigma_{i}}_{\text{interference}} + \mathbf{n}_{\ell}, 
\end{equation}
where $\mathbf{h}_{w \ell} \in \mathbb{C}^{N_{\text{DU}}}$ is the channel between the $\ell$-th AP and its serving $w$-\text{th} DU, $p_{t}^{\text{mmW}}$ is the normalized fronthaul transmission power, $\mathbf{n}_{\ell} \backsim \mathcal{N}_{\mathbb{C}}(0,\sigma^2)$ is the receiver noise at the $\ell$-\text{th} AP, and $\varsigma_{w\ell} \in \mathbb{C}$ is the signal transmitted from the $w$-\text{th} DU to AP $\ell$ with $\mathbb{E} \{|\varsigma_{w\ell}|^2\} = 1$. Nevertheless, given the relatively large distances between APs and non-serving DUs, combined with our use of beamforming, interference is assumed negligible. Consequently, the signal-to-noise ratio (SNR) at AP $\ell$ and the corresponding mmWave fronthaul link capacity can expressed as follows:
\begin{align}
    \mathbf{SNR}_{\ell} &= \frac{p_{t}^{\text{mmW}} \left | \mathbf{h}^T_{w  \ell}  \mathbf{f}_{w  \ell} \right |^2 }{\sigma^2},
\\
    R_{w \ell}^{\text{mmW}} &= BW^{\text{mmW}} \log_2(1 + \mathbf{SNR}_{\ell}).
\end{align}
\vspace{-1.7em}
\section{Fronthaul Cost Optimization Formulation}\label{Sec3}
\vspace{-2pt}
To simplify problem formulation without any loss of practicality, we assume a brownfield deployment where APs and DUs locations are known~\cite{BROWN-GREENFIELDS}. Additionally, and as discussed in Section \ref{Sec2}, we begin the formulation having all the following as inputs: APs-DUs clusters ($\mathbf{L}_w$) and distances ($d_{w\ell}$), fronthaul capacity threshold for each AP ($\psi_{\ell}$), and actual capacities for all fronthaul technologies ($R_{w\ell}^{\text{Fiber}}$, and $R_{w\ell}^{\text{mmW}}$).
\vspace{-1em}
\subsection{Objective Function}
\vspace{-3pt}
The goal is to minimize the TCO of the fronthaul network by accounting for both operational (OPEX) and capital expenditures (CAPEX) of each technology, where CAPEX is further divided into deployment costs per AP ($C_{\ell}$) and DU ($C_{w}$).

\subsubsection{Fiber-based Fronthaul}
For every AP $\ell$ that uses fiber, the associated costs include the ONU cost, installation cost, and O\&M cost for ($N_{\text{yrs}}$) number of years, and the cost of trenching and burial of fiber cables per meter ($\eta^{\text{Fiber}}$), denoted collectively as $C_{\ell}^{\text{Fiber}}$. At the DU side, the OTN cost associated with DU $w$ serving fiber-connected APs is denoted by $C_{w}^{\text{Fiber}}$. Therefore, the TCO for fiber-based fronthaul is expressed as:
\begin{equation}
\begin{aligned}
C^{\text{Fiber}} = &  \sum_{w=1}^{W} \sum_{\ell=1}^{L_w}  \left( \underbrace{C_{}^{\text{ONU}} + N_{\text{yrs}}   C_{\text{O\&M}}^{\text{Fiber}}}_{C_{\ell}^{\text{Fiber}}} + \eta^{\text{Fiber}}   d_{w \ell}\right) \\
& + \sum_{w=1}^{W} \left( \underbrace{C^{\text{OLT}} + C^{\text{MUX}} + C^{\text{Other}}}_{C_{w}^{\text{Fiber}}} \right).\\
\end{aligned}
\end{equation}
\subsubsection{mmWave-based Fronthaul}
For mmWave-based fronthaul, each AP $\ell$ is equipped with a single mmWave antenna (i.e., $N_{\text{AP}}$ = 1). The TCO per AP, including annual power consumption, installation, and O\&M costs, is denoted by $C_{\ell}^{\text{mmW}}$. On the other hand, each DU that serves mmWave APs will have a massive MIMO antenna device with $N_{\text{DU}}$ elements, with its cost denoted by $C_{w}^{\text{mmW}}$. Hence, the TCO for the mmWave-based fronthaul is given by:
\begin{equation}
\begin{aligned} 
C^{\text{mmW}} = & \sum_{w=1}^{W} \sum_{\ell=1}^{L_w}  \left( \underbrace{C_{\ell}^{\text{mmW-RX}} + N_{\text{yrs}} C_{\text{O\&M}}^{\text{mmW}}}_{C_{\ell}^{\text{mmW}}} \right) + \sum_{w=1}^{W} C_{w}^{\text{mmW}}. 
\end{aligned}
\end{equation}
By combining the TCO of all fronthaul technologies, the final joint cost objective function can be expressed as follows:

\begin{equation}
\label{finalobjective}
\begin{aligned}
& f_{\rm o}({\mathbf{u}}, {\bf z}, {\bf v}, {\bm \kappa}) = \sum_{w=1}^{W} \sum_{\ell=1}^{L_w}  \left[ u_{w \ell}  \left( C_{\ell}^{\text{Fiber}} + \eta^{\text{Fiber}} d_{w \ell} \right) \right] \\
& + \sum_{w=1}^{W} \sum_{\ell=1}^{L_w}  z_{w \ell}  C_{\ell}^{\text{mmW}} + \sum_{w=1}^{W} \left[ v_w  C_{w}^{\text{mmW}} + \kappa_w  C_{w}^{\text{Fiber}} \right],  \\
\end{aligned}
\end{equation}

where $u_{w \ell}, z_{w \ell} \in \{0, 1\}$ are the binary controlling variables for APs selecting fiber or mmWave, respectively. While $\kappa_w \in \mathbb{Z}$ is an integer variable indicating the number of fiber OTNs required at the DU side. While $v_w \in \{0, 1\}$ is a binary variable indicating if a DU $w$ has any connected mmWave APs.

\vspace{-1em}
\subsection{Constraints}\label{constraints}
\vspace{-3pt}
Effective fronthaul planning and joint optimization of multiple technologies necessitate the incorporation of practical and realistic constraints guiding the selection process of the most cost-effective technologies, and are as follows: 

\subsubsection{\textbf{General Architectural Constraints}} This category ensures robust formulation and network components association, and it includes the following constraints:

\textbf{Constraint 1-A:} Ensures that each AP $\ell$ is only connected to a single DU $w$ and selecting one fronthaul technology.
      \begin{equation}\label{first-constraint}
        \sum_{w=1}^{W} \left( u_{w \ell} +  z_{w \ell} \right) = 1 , \;\;\; \forall \ell \in \mathcal{L}. 
        \end{equation}
        
\textbf{Constraint 1-B:} Defines the binary controlling variables. \begin{equation}\label{second-constraint}
        u_{w \ell}, z_{w \ell} \in \{0, 1\}, \;\;  \forall \ell \in \mathcal{L}, \;\;\; v_{w} \in \{0, 1\}, \;\;  \forall w \in \mathcal{W}.
    \end{equation}
\textbf{Constraint 1-C:} Defines the fiber-associated controlling variable $\kappa_w$ as an integer.
    \begin{equation}\label{third-constraint}
        \kappa_w \in \mathbb{Z},  \;\;\; \forall w \in \mathcal{W}.
    \end{equation}

\subsubsection{\textbf{Technology-Specific Constraints}}
This category ensure that the unique characteristics of each technology are accurately represented, and it includes:

\textbf{Constraint 2-A} \textit{\textbf{(Fiber):}} In fiber-based fronthaul, the maximum number of APs using fiber connected to a single DU depends on the capacity ($\Theta$) of the OTN deployed at the DU side. For simplicity, we are taking the ratio option of splitters to determine the bottleneck capacity of OTNs. Assuming a single splitter supports $\Theta$ number of links (i.e., 1:$\Theta$ PON)~\cite{NO.1-Sources}, then if more than $\Theta$ APs associated with $w$-th DU are using fiber, and additional OTN must be deployed, effectively doubling the cost of $C_{w}^{\text{Fiber}}$. The constraint governing this setup is expressed by a ceiling function applied to the decision variable $\kappa_w$ as follows: 
\begin{equation}\label{fourth-constraint}
\sum_{\ell=1}^{L_w} \frac{u_{w \ell}}{\Theta}  \leq \; \kappa_w \; \leq   \sum_{\ell=1}^{L_w} \frac{u_{w \ell}}{\Theta} + 1 - \epsilon,  \;\;\; \forall w \in \mathcal{W}.
\end{equation}
\textbf{Constraint 2-B} \textit{\textbf{(mmWave):}} Similar to fiber, this constraint guarantees that the $w$-th DU will not be equipped with a mmWave antenna device unless there is at least one AP from the subset of APs, denoted by $\mathbf{L}_w$, employs mmWave technology for fronthauling. As a result, since $v_w$ is a binary variable, $v_w$ will be equal to 1 if and only if $ \sum_{\ell=1}^{L_w} z_{w \ell} \neq 0$.
    \begin{equation}\label{fifth-constraint}
    \sum_{\ell=1}^{L_w} \frac{z_{w \ell}}{L_w} \leq \; v_w \; \leq \sum_{\ell=1}^{L_w} z_{w \ell},  \;\;\; \forall w \in \mathcal{W}.
    \end{equation}

\subsubsection{\textbf{QoS Metrics Constraints}}
This category ensures that individual APs and the entire network meet performance standards, and it includes:

\textbf{Constraint 3-A:} Guarantees that the selected fronthaul technology minimizing the TCO for each AP $\ell$ is also meeting the fronthaul capacity threshold $\psi_{\ell}$. 
    \begin{equation}\label{sixth-constraint}
        u_{w \ell} R_{w \ell}^{\text{Fiber}}  + z_{w \ell}  R_{w \ell}^{\text{mmW}}    \geq  \psi_{\ell}, \;\; \forall \ell \in \mathcal{L}. 
    \end{equation}
    
\textbf{Constraint 3-B:} Guarantees that the aggregate fronthaul sum-rate of the selected fronthaul technologies meets the minimum required backhauling rate for each DU, expressed as a portion $0 < \alpha \le 1$ of the backhaul rate $W^{\text{backhaul}}$. This also ensures that we do not waste the backhauling incoming data. Hence, associated APs and selected technologies must be able to carry $\alpha$ of this traffic, where $\alpha$ approximates the packets processing ratio that takes place in DUs.
\begin{equation}\label{seventh-constraint}
         \sum_{\ell=1}^{L_w} \left( u_{w \ell}  R_{w \ell}^{\text{Fiber}} +  z_{w \ell} R_{w\ell}^{\text{mmW}} \right) \geq \alpha   W_{w}^{\text{backhaul}}, \;\forall w \in \mathcal{W}. 
\end{equation}
The optimization problem that aims to minimize the fronthaul network TCO and ensures effective performance through the selection of fronthaul technologies is presented as follows:
\begin{subequations}\label{Final-Formulation}
\begin{align}
\min_{\substack{
u_{w \ell}, \, z_{w \ell}, \\
v_{w}, \, \kappa_{w} 
}}
\quad & f_{\rm o}( \mathbf{u}, {\bf z}, {\bf v}, {\bm \kappa}),  \\
\text{subject to} \quad &
\begin{aligned}
&\eqref{first-constraint} - 
\eqref{seventh-constraint}.
\end{aligned}
\end{align}
\end{subequations}
The above formulation is a Mixed-Integer Linear Program (MILP), which is a combinatorial optimization problem characterized by the mixed combination of binary ($u_{w \ell}, z_{w \ell}, v_{w}$) and integer ($\kappa_w$) variables, alongside the linear nature of the objective function and constraints. To achieve the optimal solution of equation \ref{Final-Formulation}, we employ the branch-and-bound method~\cite{gurobi}, and the steps are outlined in Algorithm \ref{ALG1}.  
\vspace{-0.4em}
\begin{algorithm}
\caption{MILP algorithm for joint optimization of eq.~\ref{Final-Formulation}.}\label{ALG1}
\begin{algorithmic}[1]
\State \textbf{Input:} Set up the given technology-specific and general parameters values from Table \ref{input-parameters} as input to the system.
\For{realization $r = 0$}
\State {$r \gets r + 1$;}
\State \textbf{Initialization:}
\State {Uniformly distribute $L$ APs.}
\State {Deploy $W$ DUs using K-means.}
\For {$\bm{\forall w \in \mathcal{W}}$}
\State \text{Optimize $\mathbf{L}_w$ and locations of DUs.}
\State \text{Generate random $W_{w}^{\text{backhaul}}$.}
\State \text{Initialize $\mathbf{v}_{}^{(r)}$ and $\bm{\kappa}_{}^{(r)}$ $\gets$ $0$;}
\For {$\bm{\forall \ell \in \mathbf{L}_w}$}
\State \text{Calculate  $d_{w\ell}$, $R_{w\ell}^{\text{mmW}}$, and $R_{w\ell}^{\text{Fiber}}$. }
\State \text{Generate and assign $\psi_{\ell}$.}
 \State \text{Initialize $\mathbf{u}^{(r)}$ and $\mathbf{z}^{(r)}$ $\gets$ $0$; }
\EndFor
\EndFor
\State \text{Solve eq.~\ref{Final-Formulation} using branch and bound to obtain:}
\State   $\mathbf{u}_*^{(r)},  \mathbf{z}_*^{(r)}, \mathbf{v}_*^{(r)}, \text{and }  \bm{\kappa}_*^{(r)}.$
    \EndFor
\State \textbf{Output:} $\mathbf{u}, \mathbf{z}, \mathbf{v},$ $\bm{\kappa},$ $C^{\text{Fiber}},$ $C^{\text{mmW}},$ and $f_{\rm o}({\bf u}, {\bf z}, {\bf v}, {\bm \kappa}).$
\end{algorithmic}
\end{algorithm}
\begin{table}[t]
\renewcommand{\arraystretch}{1.2}
\centering
\caption{Input system parameters values used in simulation.}
\vspace{-0.2em}
\label{input-parameters}
\begin{tabular}{|c|c||c|c|}
\hline
\multicolumn{2}{|c||}{\textbf{mmWave}} & \multicolumn{2}{|c||}{\textbf{Fiber}} \\
\hline
\hline
{$C_{w}^{\text{mmW}}$} &  \$34,500 & {$\eta^{\text{Fiber}}$} & \$26/m \\
\hline
{$C_{\ell}^{\text{mmW-RX}}$} & \$6,000 & {$C_{}^{\text{ONU}}$} & \$6,502 \\
\hline
{$C_{\text{O\&M}}^{\text{mmW}}$} &  \$13,000 & {$C_w^{\text{Fiber}}$} & \$61,727  \\
\hline
{$N_{\text{AP}}$}, {$N_{\text{DU}}$} & {1}, {128} & {$C_{\text{O\&M}}^{\text{Fiber}}$} & \$2,285 \\
\hline 
{$q$}, {$p_{t}^{\text{mmW}}$} & {6}, {120 W} & {$\Theta$} & {16} \\
\hline
{$\text{BW}^{\text{mmW}}$, $f_{c}$} & 800 MHz, 28 GHz & {$R_{}^{\text{Fiber}}$} & {10 Gbps}  \\
\hline
\hline
\multicolumn{4}{|c|}{\textbf{General}} \\
\hline
\hline 
{$C_{\text{DU}}^{\text{pool}}$} & \$91,035 &  $\mathbf{X_{}^{\text{min}}}, \mathbf{X_{}^{\text{max}}}$  &  {0.1, 10} Gbps  \\
\hline
\text{Area} & {2 km $\times$ 2 km} & {$L$} & {200} \\
\hline
{$N_{\text{yrs}}$} & {1 year} & $\mathbf{\alpha}$ & 0.7 \\
\hline
{$L_w$}, {$\psi_{\ell}$} & \textbf{\textit{Variables}} & {$W$}, {$W_{w}^{\text{backhaul}}$} & \textbf{\textit{Variables}} \\
\hline
\end{tabular}
\vspace{-1.5em}
\end{table}
\vspace{-1.8em}
\section{Numerical Results and Discussions}\label{Sec4}
\vspace{-3pt}
The fronthaul technologies selection and optimized cost values are evaluated with respect to two key variables: \textbf{(1)} Varying the number of DUs (${W}$), impacting distances between DUs and APs. \textbf{(2)} Traffic-aware fronthaul capacity thresholds ($\psi_{\ell}$), where increasing the number of high-traffic hotspots make the thresholds constraint more demanding. Other variables listed in Table \ref{input-parameters} also influence outcomes and are varied in different realizations to demonstrate the robustness of our approach. The cost values of all equipment in Table \ref{input-parameters} are mainly estimated from average values reported in the US Federal Communications Commission (FCC) document~\cite{FCC-UScosts}.
\begin{figure}[t]
    \centering
    \begin{subfigure}[b]{0.48\columnwidth}
        \centering
        \includegraphics[width=\textwidth, height=1.37in]{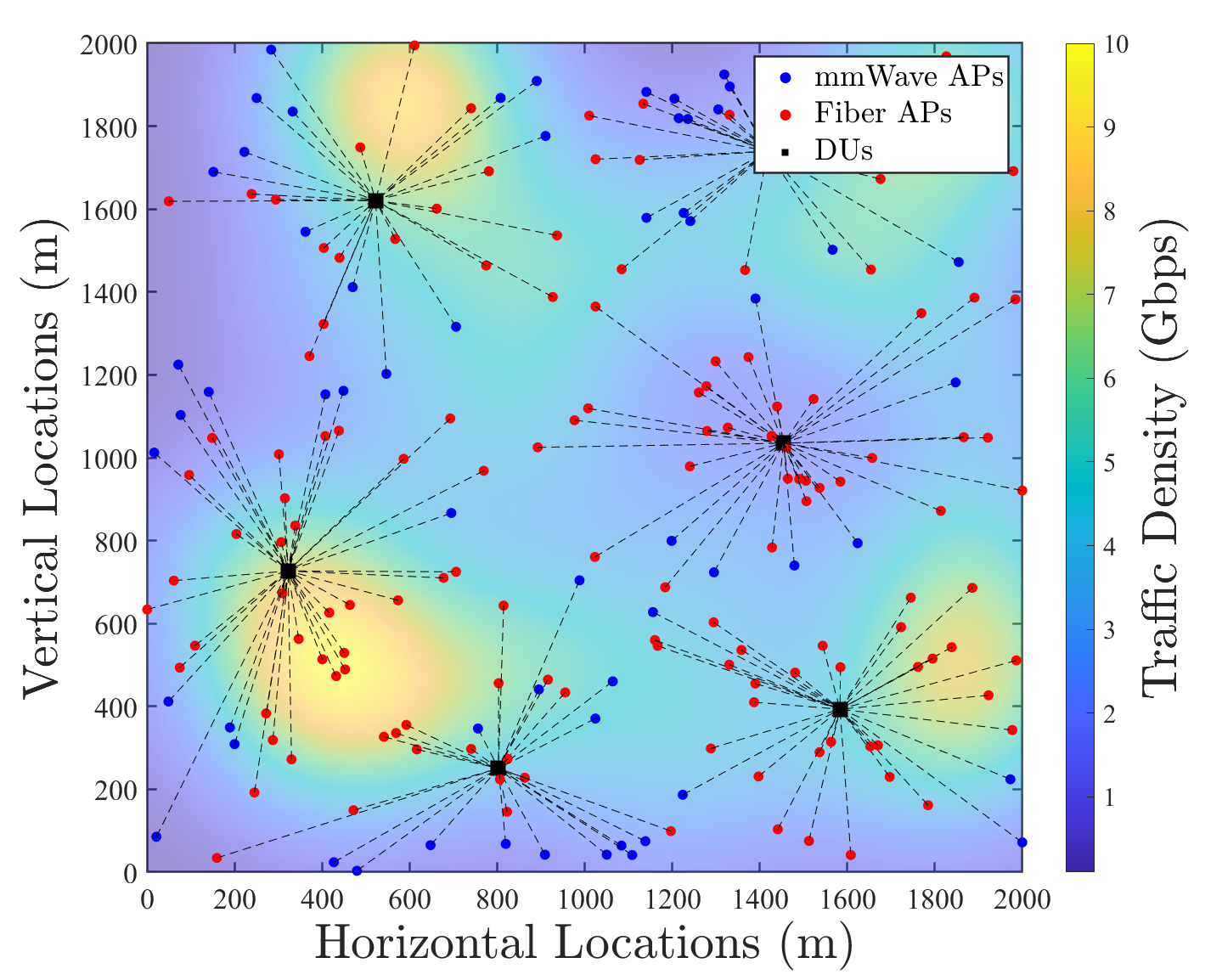}
        \caption{6 DUs with medium-traffic.}
        \label{Sample1}
    \end{subfigure}
    \hfill
    \begin{subfigure}[b]{0.48\columnwidth}
        \centering
        \includegraphics[width=\textwidth, height=1.37in]{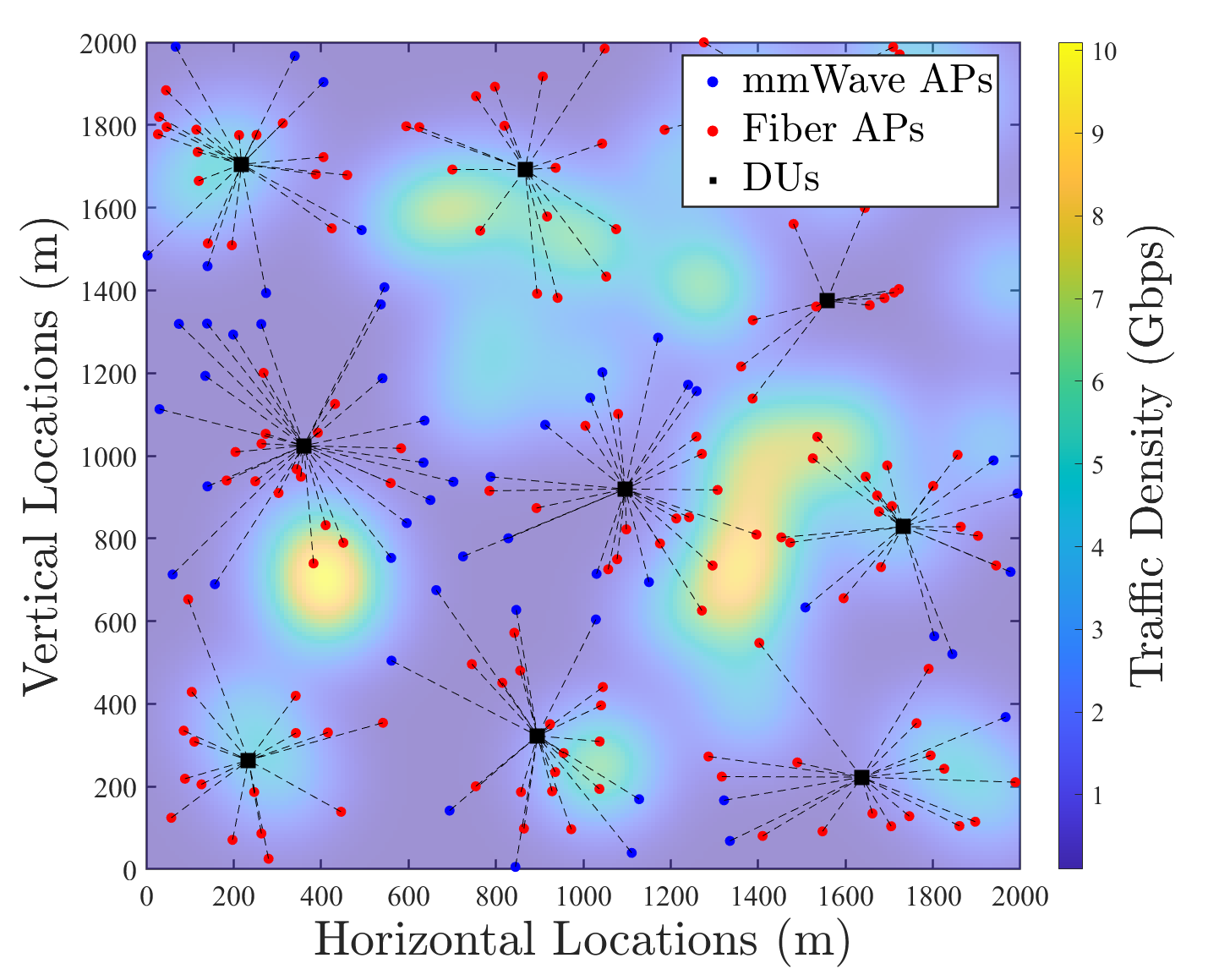}
        \caption{10 DUs with low-traffic.}
        \label{Sample2}
    \end{subfigure}
    \vspace{-0.5em}
    \caption{Network samples with optimized mixed-technologies selection for different traffic distributions and numbers of DUs.}
    \label{fig2}
    \vspace{-1.6em}
\end{figure}
We assess the performance of our optimization framework by comparing it against three benchmarks: \textbf{(a)} \textbf{Standard all-Fiber fronthaul network}, serving as a benchmark for the most reliable performance, meeting all constraints \eqref{first-constraint} - 
\eqref{seventh-constraint}, but potentially at a high TCO. \textbf{(b)} \textbf{Suboptimal all-mmWave fronthaul network}, representing an economically appealing benchmark, but falls short of meeting the traffic demands of all DUs ($W_{w}^{\text{backhaul}}$) and APs ($\psi_{\ell}$). Lastly, \textbf{(c)} \textbf{Heuristic method for hybrid deployment}, balancing cost and capacity but may not achieve optimal TCO. In this method, mmWave is initially assigned to all APs, and if the traffic demand ($\psi_{\ell}$) for AP $\ell$ surpasses its mmWave capacity $R_{w\ell}^{\text{mmW}}$, it is switched to fiber. Subsequently, we iteratively verify if constraint \ref{seventh-constraint} is satisfied for each DU $w$. If not, then a randomly selected mmWave-based AP is transitioned to fiber until the constraint is met. 
\begin{figure}[t]
    \centering
    \vspace{-0.5em}
    \begin{subfigure}[b]{0.48\textwidth}
        \centering
        \includegraphics[width=\textwidth, height=1.27in]{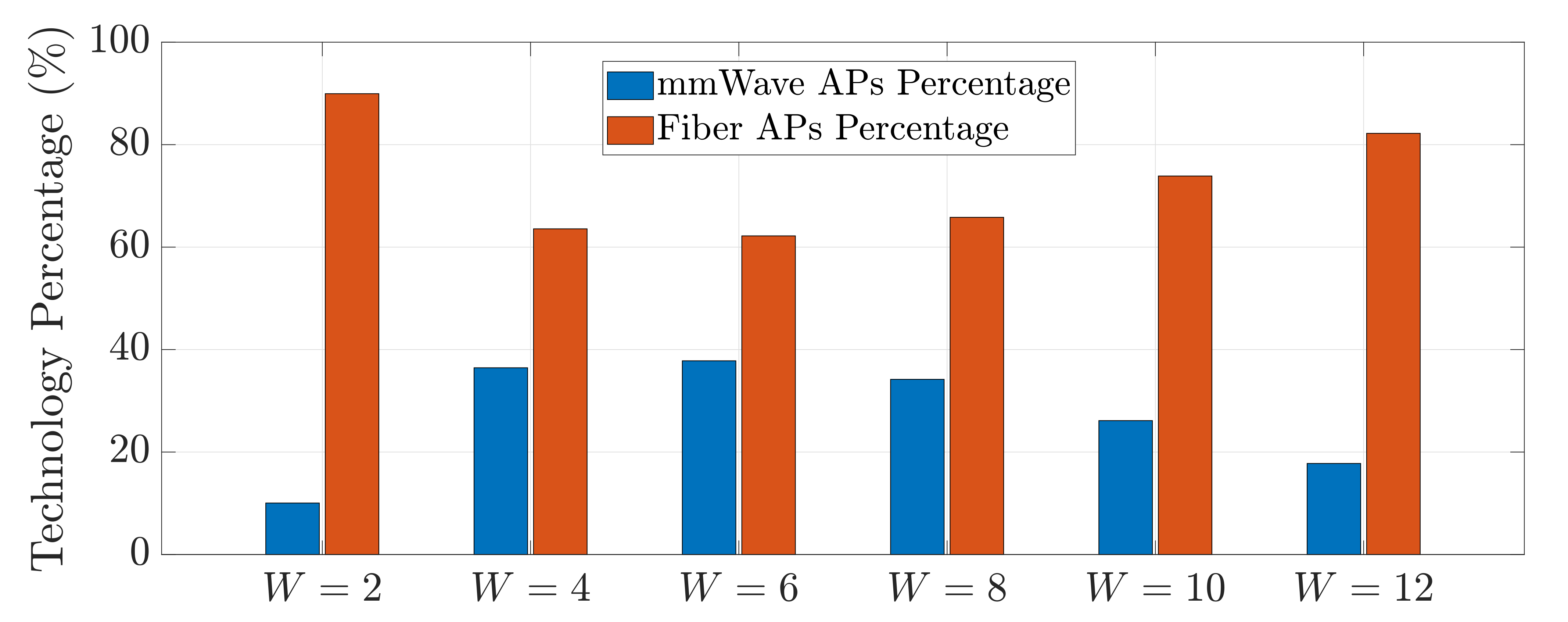}
        \caption{Average percentages of technologies selection.}
        \label{fig3}
    \end{subfigure}
    \hfill
    \begin{subfigure}[b]{0.48\textwidth}
        \centering
        \includegraphics[width=\textwidth, height=1.34in]{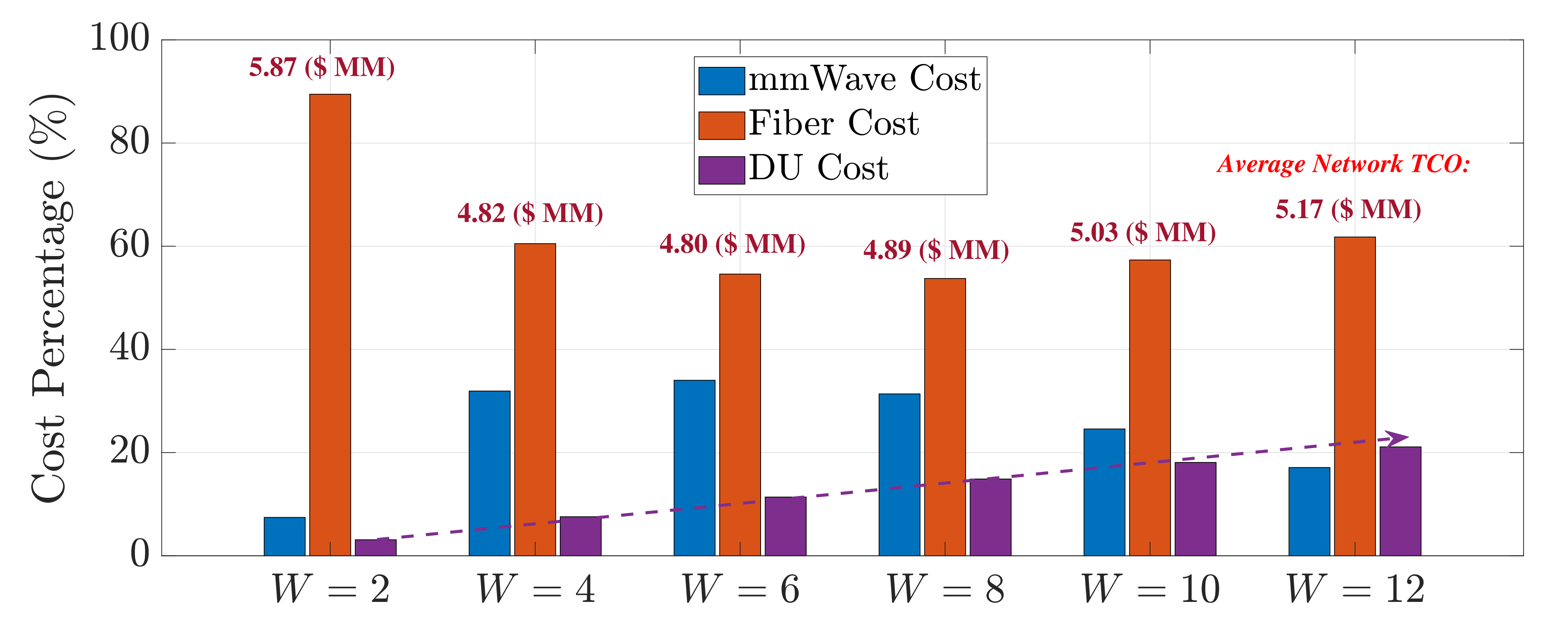}
        \caption{Average network TCO and technologies cost percentages.}
        \label{fig4}
    \end{subfigure}
    \vspace{-0.5em}
    \caption{Comparison of fronthaul technologies and network TCO for different numbers of deployed DUs.}
    \label{fig_combined}
    \vspace{-1.8em}
\end{figure}
\begin{figure*}[t]
    \centering
    \begin{subfigure}[b]{0.31\textwidth}
        \centering
        \includegraphics[width=\textwidth, height=1.55in]{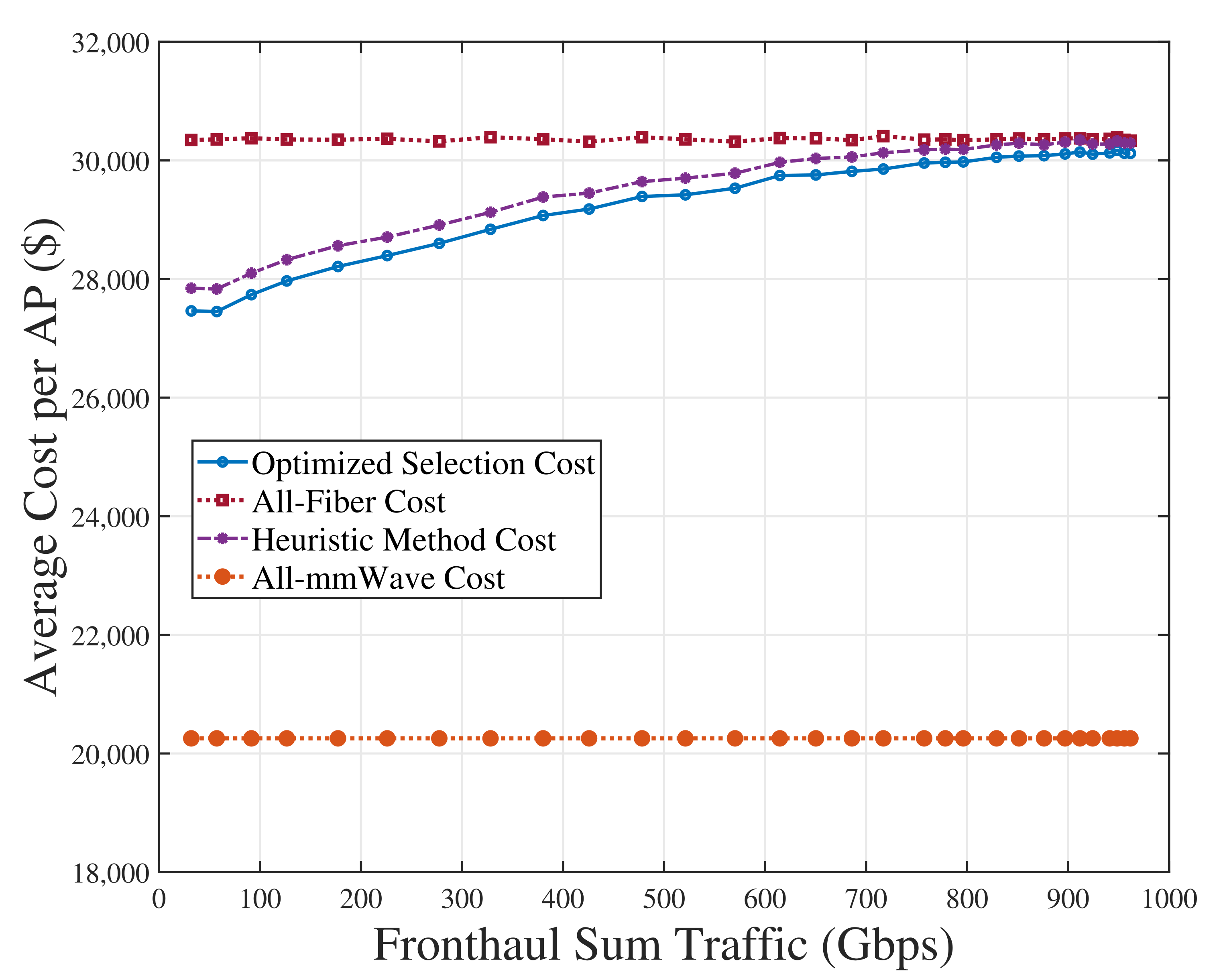}
        \caption{TCO per AP for \( W = 2 \).}
        \label{cost1}
    \end{subfigure}
    \hfill
    \begin{subfigure}[b]{0.31\textwidth}
        \centering
        \includegraphics[width=\textwidth, height=1.55in]{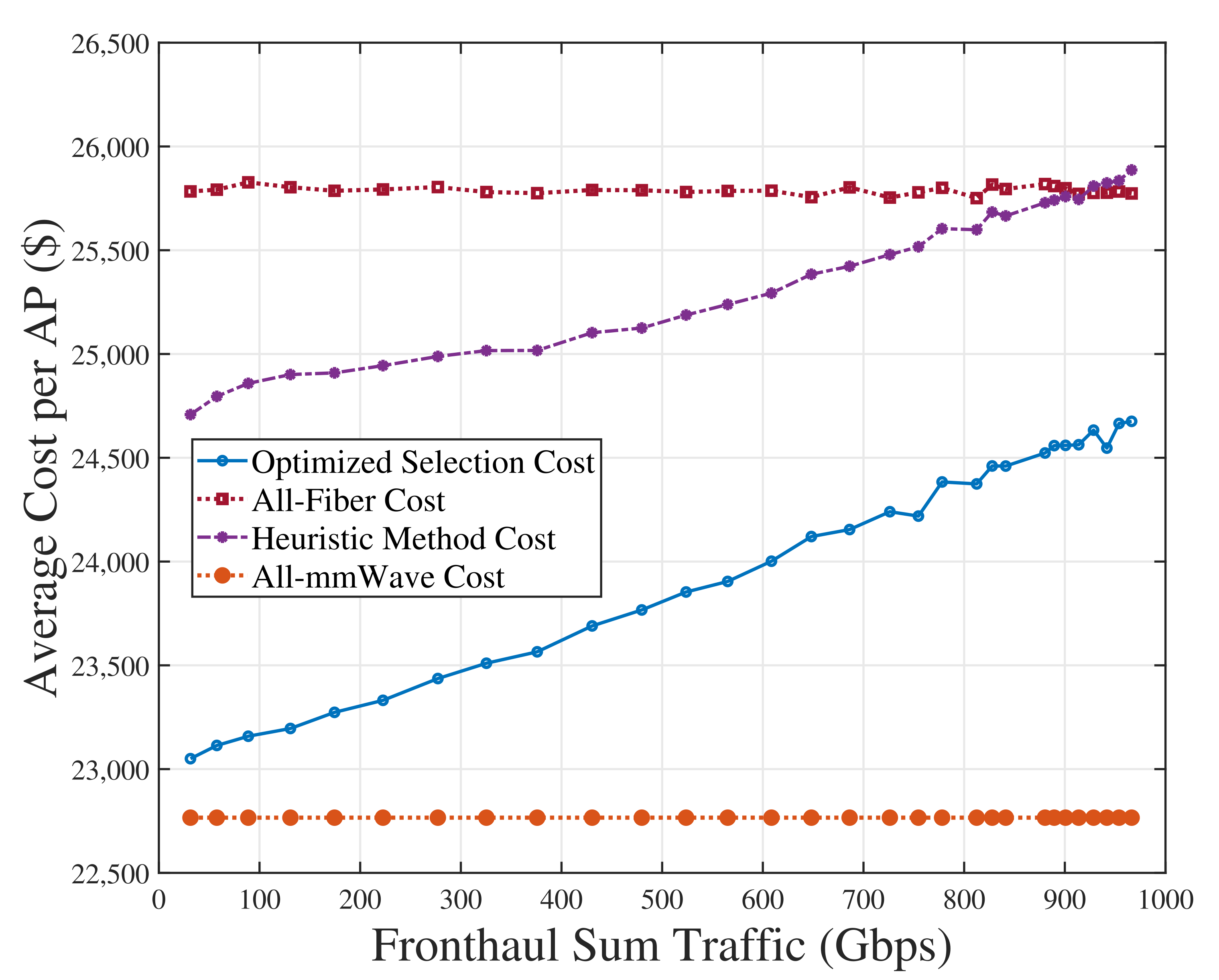}
        \caption{TCO per AP for \( W = 6 \).}
        \label{cost2}
    \end{subfigure}
    \hfill
    \begin{subfigure}[b]{0.31\textwidth}
        \centering
        \includegraphics[width=\textwidth, height=1.55in]{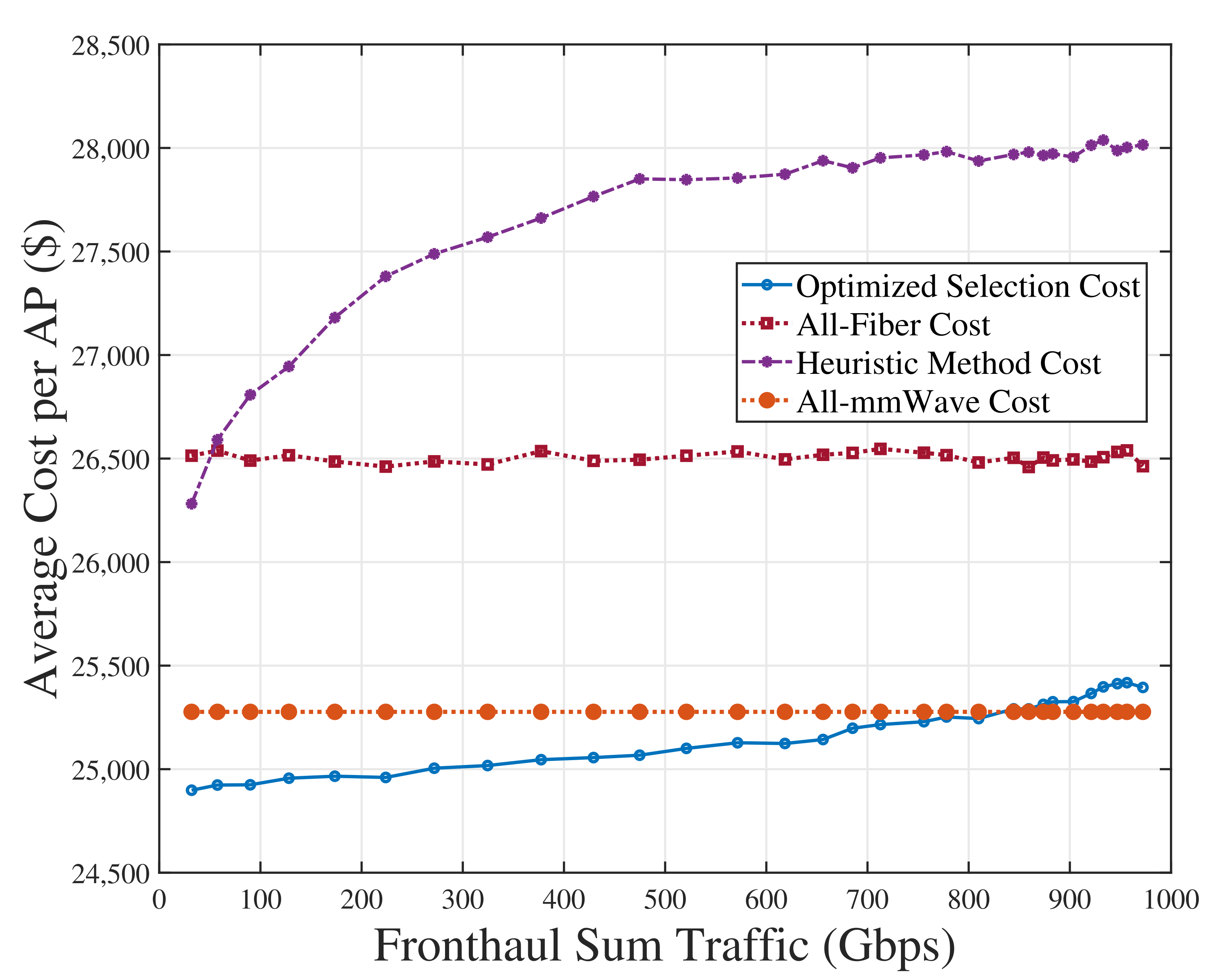}
        \caption{TCO per AP for \( W = 10 \).}
        \label{cost3}
    \end{subfigure}
    \vspace{-0.5em}
    \caption{Average TCO per AP vs. fronthaul sum traffic thresholds across different numbers of deployed DUs (\( W \)) and benchmarks.}
    \label{fig5}
    \vspace{-0.8em}
\end{figure*}
\begin{figure*}[t]
    \centering
    \subfloat[Network surplus capacity for $W$ = 2.\label{surplus1}]{
        \includegraphics[width=0.32\textwidth, height=1.55in]{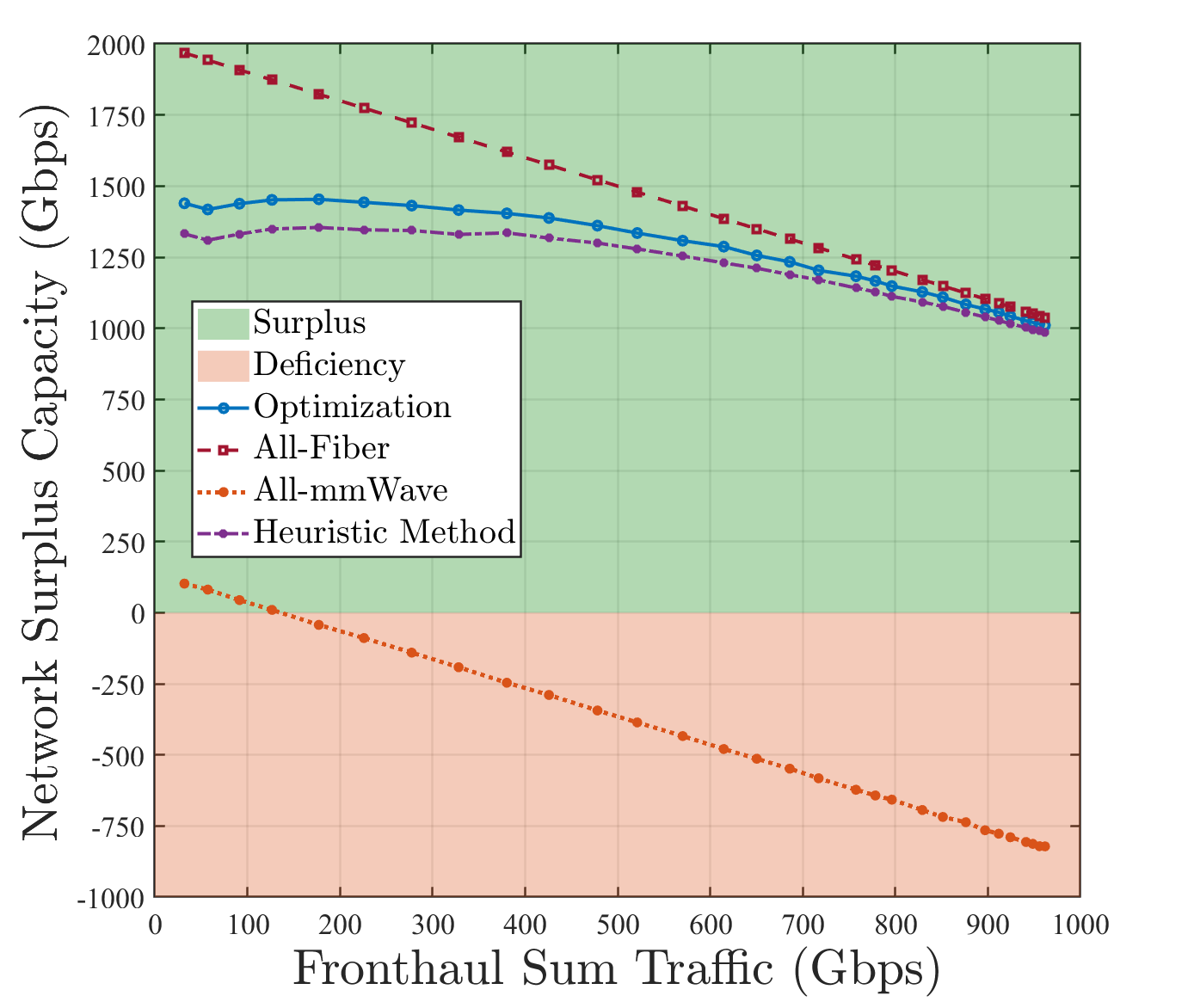}}
    \hfill    
    \subfloat[Network surplus capacity for $W$ = 6.\label{surplus2}]{
        \includegraphics[width=0.32\textwidth, height=1.55in]{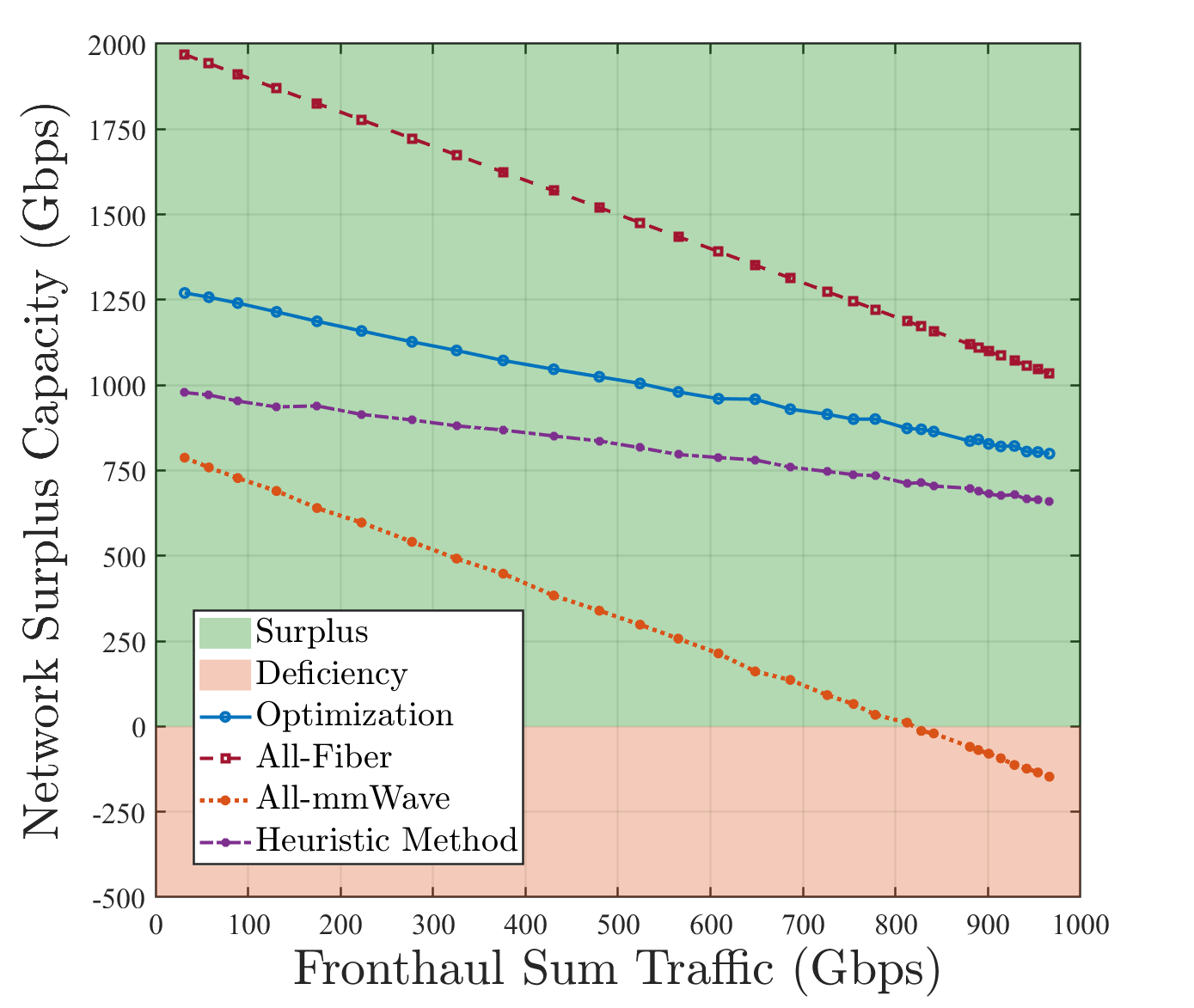}}
    \hfill
    \subfloat[Network surplus capacity for $W$ = 10.\label{surplus3}]{
        \includegraphics[width=0.32\textwidth, height=1.55in]{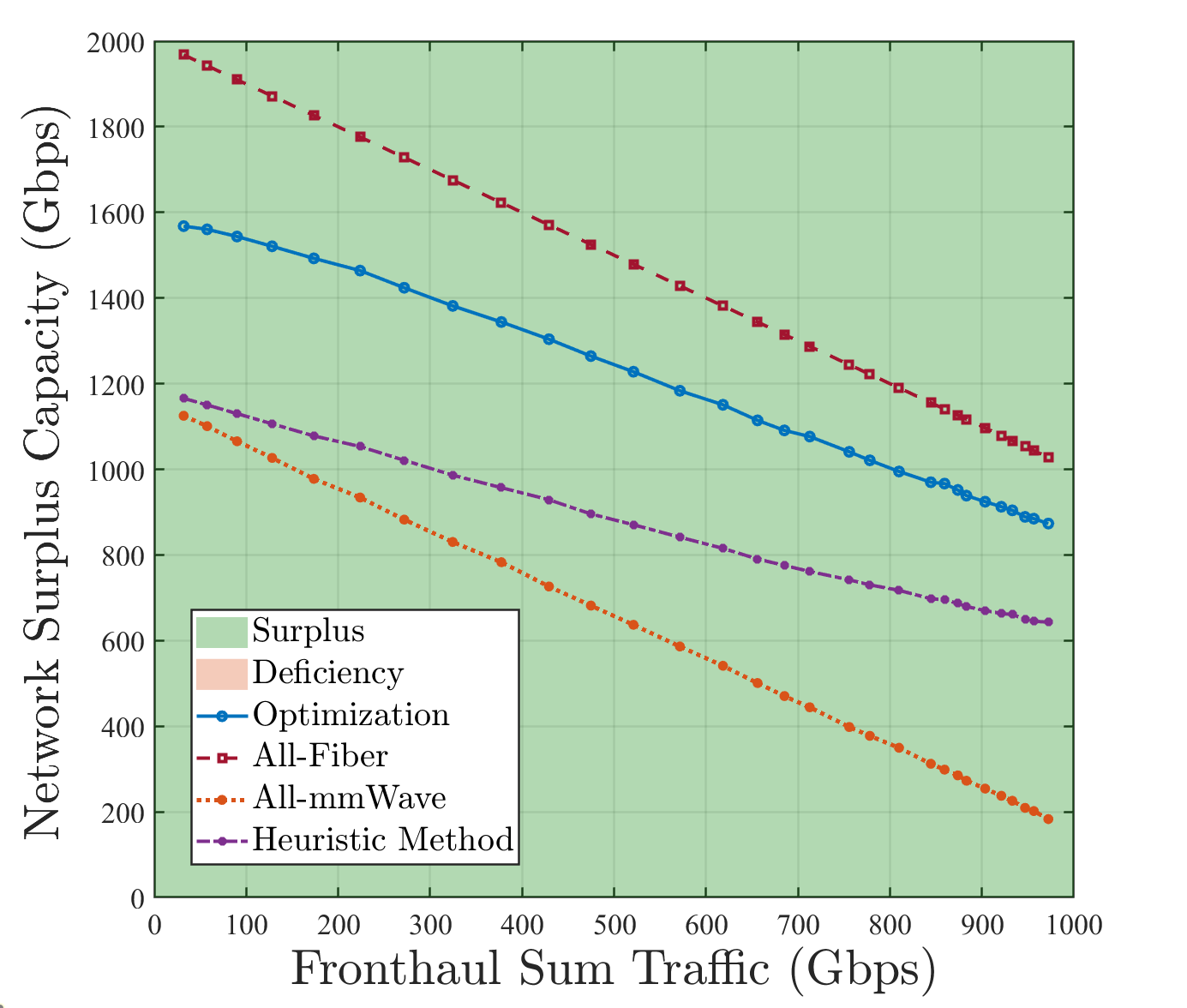}}
         \vspace{-0.5em}
    \caption{\centering Network surplus capacity vs. fronthaul sum traffic across different number of deployed DUs ($W$) and benchmarks.}
    \label{fig11}
    \vspace{-1.6em}
\end{figure*}

\vspace{-18pt}
\subsection{Fronthaul Technologies Selection and Traffic}
\vspace{-3pt}
Figures \ref{Sample1}-\ref{Sample2} show samples of hybrid fronthaul technology selection, revealing a preference for mmWave in low-traffic areas with fewer DUs, while fiber is more favored for APs situated in close proximity to DUs due to lower installation costs. Also, mmWave is often chosen for distant APs in low-traffic areas if constraints are met. Contrary to widely held assumptions, we observe that as $W$ increases and AP-DU distances decrease, fiber becomes more cost-effective, even in low-traffic regions. Additionally, the optimization framework favors exploiting the utilization of DU-associated equipment for each selected fronthaul technology, maximize the number of connections using the same technology to reduce infrastructural redundancy. Nevertheless, fronthaul technology selections are highly contingent to the traffic distribution. To provide a broader perspective, Figure \ref{fig3} presents selection percentages averaged using Monte Carlo simulations across various fronthaul traffic densities and DU values (${W}$). We see that in sparse DUs deployment ($W = 2$), fiber is dominant despite its higher cost, due to its high capacity over long-distances compared to mmWave. As $W$ increases to $4$-$8$, distances between DUs and APs become more manageable, and the combination of fiber and mmWave emerges as viable strategy to construct a cost-effective fronthaul network. With higher number of DUs, shorter AP-DU distances reduce the cost of fiber deployment, making it a more favorable option.
\vspace{-14pt}
\subsection{Network Optimized TCO and Traffic}\label{TCOandTraffic}
\vspace{-3pt}
Figure \ref{fig5} illustrates the average TCO per AP for varying numbers of DUs, and hotspot densities. We observe that with small number of deployed DUs, all-fiber deployment incurs the highest cost, while the suboptimal all-mmWave scheme has the lowest cost, but it fails to meet all QoS constraints as seen in Figures \ref{surplus1} and \ref{surplus2}, and is, therefore, not a viable solution. Additionally, at a higher number of deployed DUs, the cost-effectiveness of the suboptimal all-mmWave scheme begins to diminish. Unlike common belief, this shift occurs due to the reduced AP-DU distances, which sometimes render fiber to be a more economically appealing option compared to mmWave. Also, the cost of the heuristic method tends to exceed the all-fiber scheme as $W$ and traffic increase, primarily because the network is not exploiting the DUs infrastructure effectively. Therefore, we conclude that the best strategy in UDNs involves a well-planned and diversified mix of fronthaul technologies, as exemplified by the superior performance of our optimized network. To generalize these results, Figure \ref{fig4} showcases the average network TCO in Millions of US dollars (\$ MM), and cost contribution percentages of network components, showing insights into the economic considerations associated with different deployment strategies and traffic demands. 
\vspace{-1em}
\subsection{Network Surplus Capacity}\label{surplus}
\vspace{-3pt}
The effectiveness of deployment strategies can be assessed by their surplus capacities, reflecting the difference between the total capacity provided by the deployed fronthaul network and the actual traffic demand. A positive surplus indicates that the network has excess capacity, offering room for traffic growth or accommodating unexpected surges, and a deficiency signifies that the network is operating below the traffic demand. Figure \ref{fig11} shows that while an all-fiber network offers the highest capacity, a hybrid optimized deployment consistently achieves significant surplus capacity to meet current and future demands at a lower cost, as illustrated in Figure \ref{fig5}. Conversely, all-mmWave scheme often fails to meet minimum required rates ($\psi_{\ell}$), highlighting that cost considerations should not overshadow QoS requirements to reliably meet performance targets for SPs. Lastly, the heuristic method often results in lower surplus capacity compared to the optimized network, stressing the importance of effective fronthaul planning.

\vspace{-0.8em}
\section{Conclusion}\label{Sec5}
\vspace{-4pt}

This paper presented an optimization framework for hybrid fronthaul network planning aimed at minimizing TCO for UDN deployment. The developed framework integrated fiber and mmWave links as fronthaul options, alongside key QoS metrics, to ensure robust performance in UDNs. It can also be extended to incorporate other technologies, such as Microwave or FSO, by including the relevant parameters. In addition to its applicability to support different network schemes, such as small cells. The results demonstrated the superiority of mixed technologies over single-technology deployments and the heuristic benchmark. Our findings highlighted the critical necessity of optimizing fronthaul networks to build future-proof networks that can adapt to evolving traffic demands while ensuring cost-effectiveness and adherence to stringent QoS requirements, providing instrumental insights for SPs during fronthaul planning for next-generation wireless networks.

\vspace{-0.05em}
\bibliographystyle{IEEEtran}
\vspace{-5pt}
\bibliography{IEEEabrv}

\end{document}